\documentclass[aps,preprint]{revtex4}
\usepackage{graphicx,amsmath,multirow,amssymb}

\draft
\begin{document}
\author{Bin Wang$^{a}$}
\address{$^{a}$ Department of Physics, Southwest Petroleum University, Nanchong 637001, P. R. China}
\title{Layers of quark-gluon plasma induced by chiral magnetic effect and the separation of $u,d$ quarks' critical end points }
\begin{abstract}
In this work we show that the chiral magnetic effect (CME) could cause the quark-gluon plasma (QGP) approximately forming three layers along the strong magnetic field characterised by different compositions of quark chemical potentials. This phenomenon may bring new observable outcomes, for example different layers would have different CEPs and the $u,d$ quarks would have different critical end points (CEPs). Since the $s$ quark would make the up and the down layers have some 'asymmetry' on chemical potentials, there is the possibility that at some special conditions one flavor of $u,d$ quarks lies at CEP in one layer but another flavor dose not in any layer. These results may be helpful in testing the existence of CME.

\bigskip


\bigskip

\bigskip

\end{abstract}
\maketitle

As can break the P and CP symmetry in the strong interaction level, the chiral magnetic effect (CME) has got much attention in high energy physics for many years. But from the data obtained in heavy ion colliders we still can not assert the existence of this effect\cite{a1,a2}. In this paper we will show that the CME will cause the quark-gluon plasma (QGP) approximately forming three layers along the magnetic field characterised by different compositions of quark chemical potentials.
Then we study the chiral phase transition in different layers of QGP by the coupled quark-gluon Dyson-Schwinger Equations(DSEs)\cite{HW}.
In the literature the $u,d$ quarks always have the same critical end points (CEPs) \cite{phasedia} even when they have different chemical potentials\cite{a7}. This looks naturally since they strongly interact with each other and have similar very light masses.
But our calculation shows that the CEPs of $u,d$ quarks would separate and different layers of QGP would have different CEPs. If there is the CME of $s$ quark, there is the possibility that at some special conditions one flavor of $u,d$ quarks lies at CEP in one layer but another flavor dose not in any layer. The results of this work may be helpful in testing the existence of CME.

\begin{figure}
  \includegraphics[width=15cm]{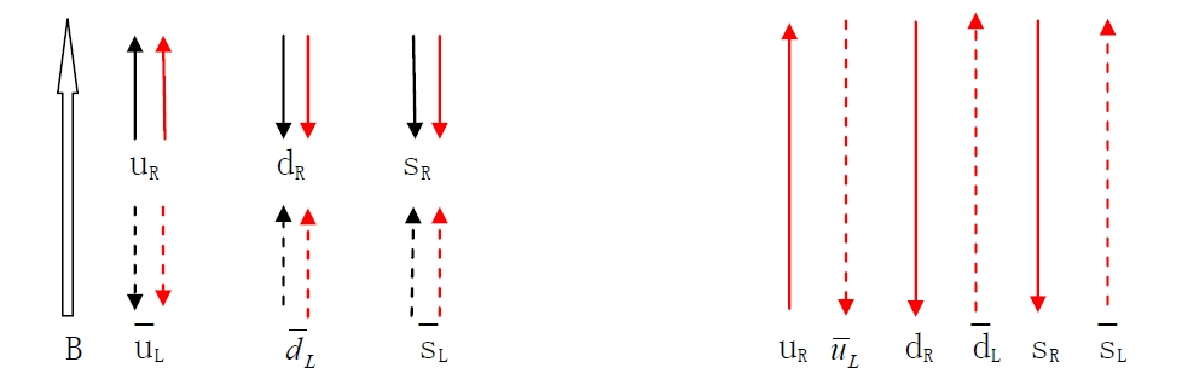}\\
  \caption{The behavior of surplus quarks and anti-quarks ($N_5>0$) in strong magnetic field.}\label{Fig1}
\end{figure}

The chiral magnetic effect accompanies the presence of net chiral charge $N_5$ in the quark gluon plasma \cite{a3},
\begin{equation}
N_5=N_R-N_L
\end{equation}
in which $N_R$/$N_L$ is the number of righthanded/lefthanded quarks. If we assume the net chiral charge is positive when the CME occurs, there would be surplus righthanded quarks ($u_R, d_R, s_R$) and righthanded anti-quarks ($\bar{u}_L, \bar{d}_L, \bar{s}_L$).
Their behaviors are plotted in the left hand of Fig.\ref{Fig1}. The magnetic field is represented by the bold vacant arrow, each quark or anti-quark is represented by two arrows, in which the black one means the spin direction and the red one means the momentum direction. With a strong magnetic field, the directions of the magnetic momentum of quarks or anti-quarks tend to be consistent with the magnetic field statistically, so that the spin of particles (or antiparticles) with positive charges will be parallel with magnetic field and the spin of particles (or antiparticles) with negative charges will be anti-parallel with the magnetic field.
The resulting moving directions of all quarks and anti-quarks are plotted again in the righthand of Fig.\ref{Fig1}. We can find that the upside of the QGP tends to have redundant $u \bar{d} \bar{s}$ while the downside tends to have redundant $\bar{u} d s$. The outcome is given in Table.\ref{tab1} in which both $\mu_C^{ud}$ and $\mu_C^{s}$ are positive real numbers which represent the variances of $u,d,s$ quarks chemical potentials induced by the CME. Since the s quark is heavier and has smaller density than $u,d$ quarks, so $\mu_C^{s}$ would be smaller than $\mu_C^{ud}$. We note that no matter which direction the magnetic field is in and which handed quarks prevail, the direction with more redundant $s$ quarks is always the direction with more net negative electric charges.
\begin{table}[htbp]
\centering\caption{The three layers of QGP induced by CME. $\mu_{B}$ is the initial baryon chemical potential without CME. $\mu_C^{ud}$ and $\mu_C^{s}$ are the CME-induced chemical potentials of $u,d$ and $s$ quarks respectively. The middle layer would disappear if the CME is strong enough.}\label{tab1}
\begin{tabular}{|c|}
\hline
 $\mu_u=\mu_{B}/3+\mu_C^{ud},\quad \mu_d=\mu_{B}/3-\mu_C^{ud}, \quad \mu_s=-\mu_C^{s}$ \\
\hline
$\mu_u=\mu_d=\mu_{B}/3, \quad \mu_s=0$ \\
 \hline
$\mu_u=\mu_{B}/3-\mu_C^{ud},\quad \mu_d=\mu_{B}/3+\mu_C^{ud}, \quad \mu_s=\mu_C^{s}$ \\
 \hline
\end{tabular}
\end{table}

Now we concentrate on the chiral phase transition of different layers. At first we show that the magnetic field $B$ and the chiral chemical potential $\mu_5$ can be neglected in numerical calculation of CME although they play important role by inducing the layers of QGP. The strong magnetic field created in hadron collision would decrease steeply with time. In the neighborhood of chiral phase transition $eB$ would be no more than several $\mathrm{MeV^2}$\cite{magrev,magpha}. But references \cite{a7,a8,a9,a10,magpha,b1,b2,b3,b4,b5,b6,b7,b8,b9,b10} indicate that the magnetic field can have apparent influence on phase transition only when $eB>0.1\mathrm{GeV}^2$.
With the method given in \cite{a3} we can estimate that $\mu_5\approx2T_0$, in which $T_0$ is the temperature after heavy-ions collision which is about 350MeV in RHIC. The QGP will inflate dramatically from initialing to phase transition. Given it inflates by ten times, $\mu_5$ will be diluted to 70MeV \cite{rhic}.
In addition $\mu_5$ would be dramatically suppressed by the quark mass which becomes big in the neighborhood of CEP \cite{a3}. So that $\mu_5$ would be much smaller than $50\mathrm{MeV}$ at last. The calculations by the DSEs with different model gluon propagators indicate that the influence of $\mu_5$ on the position of CEPs will be small if $\mu_5$ is smaller than 50MeV \cite{c1,c11}, the work based on PNJL model shows similar property \cite{c12}.
The outcomes based on other nonperturbative tools also indicate that the $\mu_5$ given above can not cause significant influence on chiral phase transitions \cite{c2,c3,c4,c5,c6,c7,c8,c9,c10}.
In following calculations, only temperature $T$ and chemical potential $\mu$ are apparently included. The influences of $B$ and $\mu_5$ are implicitly included in the compositions of chemical potentials. We will use the coupled Dyson-Schwinger Equations(DSEs) of the quark and gluon propagators \cite{HW} to estimate the CEPs.
At finite temperature and chemical potential the DSE of quark propagator can be written as\cite{roberts2000,Fischer2019}
\begin{equation}\label{gapeq}
G_a^{-1}(\widetilde{p_k})=i\gamma\cdot \widetilde{p_k}+m_a+\frac{4}{3}T\sum_{n=-\infty}^{+\infty}\int\frac{d^3q}{(2\pi)^3}g^2D_{\mu\nu}(\widetilde{p_k}-\widetilde{q_n})\gamma_{\mu}G_a(\widetilde{q_n})
\Gamma_{\nu}(\widetilde{q_n},\widetilde{p_k}).
\end{equation}
in which $a=u,d,s, \widetilde{p_k}=(\widetilde{\omega_k},\vec{p})=[(2k+1)\pi T+i\mu_a,\vec{p}]$,  $\widetilde{\omega_k}=(2k+1)\pi T+i\mu_a$, $m$ is the current quark mass ($m_u=m_d=5\mathrm{MeV}, m_s=150\mathrm{MeV}$), $G(\widetilde{p_k})$ is the full quark propagator, $D_{\mu\nu}(\widetilde{p_k}-\widetilde{q_n})$ is the full gluon propagator and $\Gamma_{\nu}(\widetilde{q_n},\widetilde{p_k})$ is the full quark-gluon vertex.
The inverse of the quark propagator $G^{-1}(\widetilde{p_k})$ can be decomposed as
\begin{equation}
G^{-1}(\widetilde{p_k})=i\vec{\gamma}\cdot\vec{p}A(\widetilde{p_k}^2)+i\gamma_4\widetilde{\omega_k} C(\widetilde{p_k}^2)+B(\widetilde{p_k}^2).
\end{equation}
The quark-gluon vertex and the gluon propagator are generally given by model in practice. At zero chemical potential one generally used truncation scheme is the bare approximation of the quark-gluon vertex, i.e. $\Gamma_\mu\rightarrow\gamma_\mu$, and the Qin-Chang model gluon propagator which is in Landau gauge\cite{QC}
\begin{equation}
g^2D^{QC}_{\mu\nu}(Q;\mu_{u,d,s}=0)=\frac{4\pi^2}{\omega^4}D_0e^{-{Q^2}/{\omega^2}}(\delta_{\mu\nu}-\frac{Q_{\mu}Q_{\nu}}{Q^2}),\label{qc}
\end{equation}
in which $Q=(\vec{p}-\vec{q},\tilde{\omega_k}-\tilde{\omega_n})$, ${\omega}D_0 = (0.80\mathrm{GeV})^3$, $\omega=0.548$. This truncation scheme could give reasonable results in calculating hadronic properties\cite{QC}. With this truncation scheme and the chosen parameters the critical temperature of nuclear matter will be given as 150 MeV, which is consistent with Lattice QCD result \cite{Tc1,Tc2}.
At finite chemical potential, the gluon propagator should be changed by \cite{HW}
\begin{equation}\label{t2}
\tilde{g}^{-2}\tilde{D}^{-1}_{\mu\nu}(Q;\mu_{u,d,s})-\tilde{g}^{-2}\tilde{D}^{-1}_{\mu\nu}(Q;0)=\eta\sum_{a=u,d,s}[{\Pi}^a_{\mu\nu}(Q;\mu_a)-{\Pi}^a_{\mu\nu}(Q;0)].
\end{equation}
The terms with tilde means they are model given. ${\Pi}^a (a=u,d,s)$ means the lowest order quark loop divided by $g^2$. $\eta$ is the correction factor which is necessary to keep the previous equation array having reasonable solutions, in this paper we use $\eta=0.02$. The presence of such correction factor is caused by the bare vertex approximation and the neglecting of gluon and ghost loops.
\begin{figure}
  \includegraphics[width=15cm]{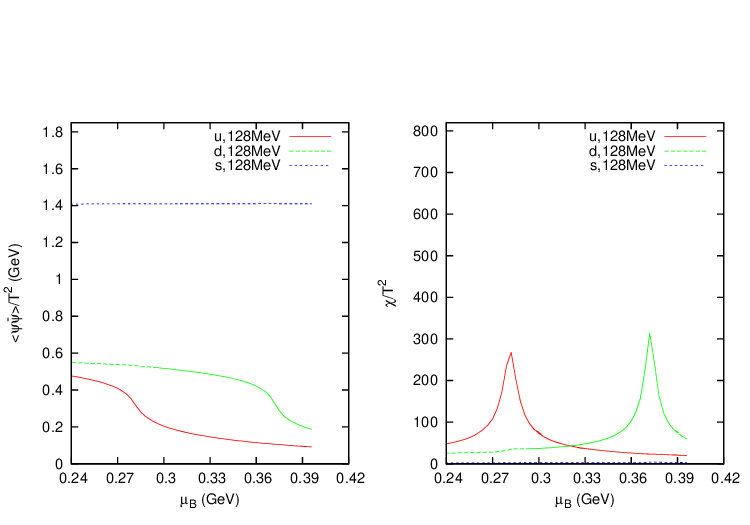}\\
  \caption{The quark condensate and chiral susceptibility for $u,d,s$ quarks when the temperature at 128MeV. (For the up layer of Table.\ref{tab1}, with $\mu_{C}^{ud}=15\mathrm{MeV}, \mu_{C}^{s}=10\mathrm{MeV}$).}\label{qq128}
\end{figure}
\begin{figure}
  \includegraphics[width=15cm]{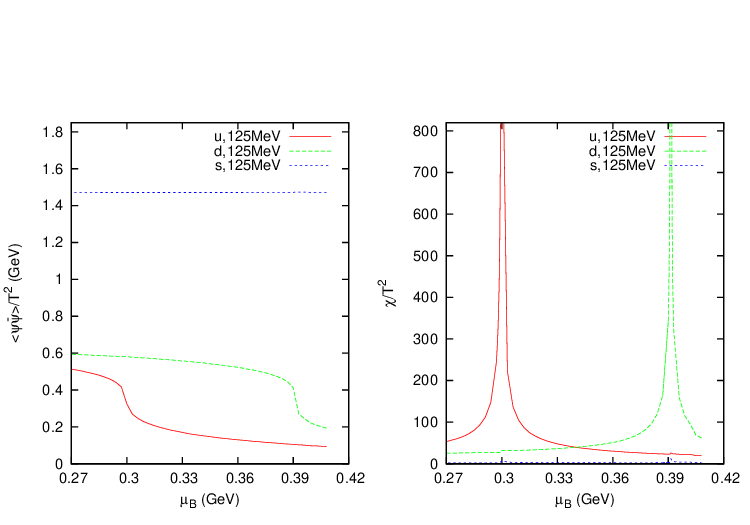}\\
  \caption{The quark condensate and chiral susceptibility for $u,d,s$ quarks when the temperature at 125MeV. (For the up layer of Table.\ref{tab1}, with $\mu_{C}^{ud}=15\mathrm{MeV}, \mu_{C}^{s}=10\mathrm{MeV}$).}\label{qq125}
\end{figure}
\begin{figure}
  \includegraphics[width=15cm]{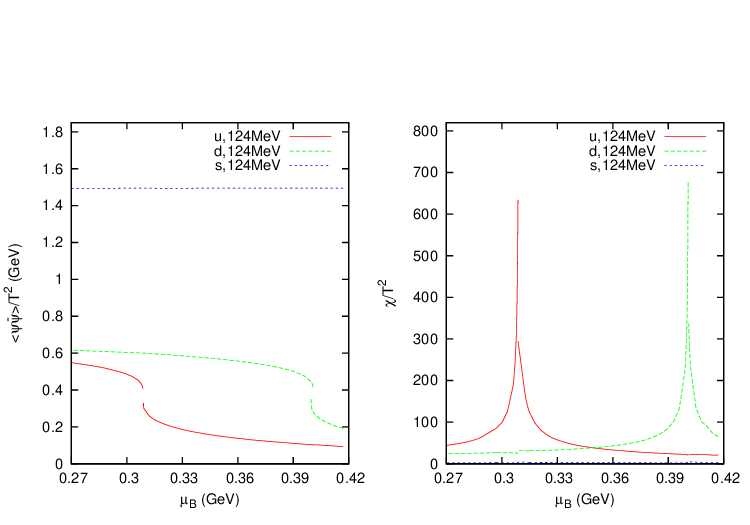}\\
  \caption{The quark condensate and chiral susceptibility for $u,d,s$ quarks when the temperature at 124MeV. (For the up layer of Table.\ref{tab1}, with $\mu_{C}^{ud}=15\mathrm{MeV}, \mu_{C}^{s}=10\mathrm{MeV}$).}\label{qq124}
\end{figure}

With the previous equations we can calculate the propagators of $u,d,s$ quarks. Then we can use them to study the chiral phase transition by calculating the quark condensate and the chiral susceptibility. The quark condensate and the chiral susceptibility at finite temperature can be written as
\begin{eqnarray}
\label{qq}
-\langle\overline{\psi}\psi\rangle_a&=&N_cT\sum_{k=-\infty}^{+\infty}\int\frac{d^3q}{(2\pi)^3}tr[G_a(p_k)], \\
\chi_a&=&(-)\frac{\partial}{\partial m_a}<\overline{\psi}\psi>_a,\quad  (a=u,d,s)
\end{eqnarray}
for each flavour of quark, in which $N_c$ is the color factor, the trace is over Dirac indices.
When $T>T_{CEP}$ both curves of quark condensate and chiral susceptibility along with chemical potential are continuous and the quark condensate curve is steepest at the critical point while the chiral susceptibility curve has a finite peak. As the temperature decreases to $T_{CEP}$ the quark condensate curve becomes steeper and steeper and the peak of chiral susceptibility curve becomes higher and higher (toward to infinity at last).  When $T<T_{CEP}$ both curves of quark condensate and chiral susceptibility are discontinuous and there is a chemical potential segment in which both curves have two values (one value corresponding to Nambu solution, another corresponding to Wigner solution). As the temperature increases from a lower value to $T_{CEP}$ such chemical potential segment would become smaller and smaller and at last shrinks to a point at $(\mu_{CEP},T_{CEP})$.
\begin{table}[htbp]
\centering\caption{The CEPs of $u,d$ quarks in $\mu_B-T$ plane in the three layers of QGP induced by CME (with the assumption $\mu_{C}^{ud}=15\mathrm{MeV}, \mu_{C}^{s}=10\mathrm{MeV}$).}\label{tab2}
\begin{tabular}{|c|c|c|}
\hline
{Up}&   $\mathrm{CEP_u}: (\mu_{B},T)=(299.7,125)\mathrm{MeV}$ & $\mathrm{CEP_d}: (\mu_{B},T)=(391.5,125)\mathrm{MeV}$ \\
\hline
{Mid}&   $\mathrm{CEP_{u}}: (\mu_{B},T)=(334.5,126)\mathrm{MeV}$ &   $\mathrm{CEP_{d}}: (\mu_{B},T)=(334.5,126)\mathrm{MeV}$ \\
 \hline
{Down}&  $\mathrm{CEP_u}: (\mu_{B},T)=(391.2,125)\mathrm{MeV}$ & $\mathrm{CEP_d}: (\mu_{B},T)=(299.7,125)\mathrm{MeV}$ \\
 \hline
\end{tabular}
\end{table}
\begin{table}[htbp]
\centering\caption{The CEPs of $u,d$ quarks in $\mu_C^{ud}-T$ plane in three CME-induced layers of QGP when $\mu_B=0$ (with the assumption $\mu_{C}^{s}=2/3 \mu_{C}^{ud}$).}\label{tab3}
\begin{tabular}{|c|c|c|}
\hline
{Up}&   $\mathrm{CEP_u}: (\mu_C^{ud},T)=(117.2,125)\mathrm{MeV}$ & $\mathrm{CEP_d}: (\mu_C^{ud},T)=(115.9,125)\mathrm{MeV}$ \\
\hline
{Mid}&   $\mathrm{No}\quad \mathrm{CEP}$ &   $\mathrm{No}\quad \mathrm{CEP}$ \\
 \hline
{Down}&  $\mathrm{CEP_u}: (\mu_C^{ud},T)=(115.9,125)\mathrm{MeV}$ & $\mathrm{CEP_d}: (\mu_C^{ud},T)=(117.2,125)\mathrm{MeV}$ \\
 \hline
\end{tabular}
\end{table}

To study the effect of the CME-induced chemical potential on the chiral phase transition we first assume $\mu_{C}^{ud}=15\mathrm{MeV}, \mu_{C}^{s}=10\mathrm{MeV}$. For the up layer of QGP in Table.\ref{tab1}, the curves of quark condensate and chiral susceptibility along with $\mu_B$ are shown in Fig.\ref{qq128},Fig.\ref{qq125} and Fig.\ref{qq124}. We can find that the $u,d$ quarks have different CEPs, for u quark $(\mu_{B},T)_{CEP}=(299.7,125)\mathrm{MeV}$, for d quark $(\mu_{B},T)_{CEP}=(391.5,125)\mathrm{MeV}$. With the same method we can determine the CEPs for other two layers. The results are given in Table.\ref{tab2}. We can see that in the middle layer the $u,d$ quarks have the same CEPs, but in other two layers the $u,d$ quarks have different CEPs. The difference of the $\mathrm{CEP_d}$ in up layer and the $\mathrm{CEP_u}$ in down layer is caused by $\mu_C^s$. Although this difference is small, but its value might increase (or decrease) if the parameter or the truncation scheme changes. The $\mathrm{CEP_u}$ in up layer is the same to the $\mathrm{CEP_d}$ in down layer, their difference might be too small to be detected here (since the step of T/$\mu$ is MeV/0.1MeV in our calculation).
One important property in previous results is that at some special conditions one flavor of $u,d$ quarks lies at CEP in one layer but another flavor dose not in any layer. This could cause measurable results, for example, the hadron number relating to $u\,\bar{u}$ might be different from that relating to $d\,\bar{d}$. This property would maintain even if we include magnetic field and chiral charge, since it is caused by the asymmetry of chemical potential in the up and down layers of QGP. (Since the phase diagrams of matter and anti-matter are similar and the $u,d$ quarks play equal role in chiral phase transition, so the up layer and down layer can be viewed as symmetric in chemical potential when $\mu_C^s=0$, but can be viewed as asymmetric when $\mu_C^s\neq0$.)

Besides making the $u,d$ quarks have different CEPs in $\mu_B-T$ plane, the CME also could strengthen the intensity of chiral phase transition from crossover to first order when $\mu_{B}=0$. To show this property, we assume $\mu_{C}^{s}=2/3 \mu_{C}^{ud}$, then with the same procedure as before the CEPs in the $\mu_{C}^{ud}$-T space can be determined, they are given in Table.\ref{tab3}. We can see that the $u,d$ quarks in the same side have different CEPs, but the u (d) quark in upside and the d(u) quark in downside have the same CEPs.

Last but not least, why dose the $u,d$ quark can have different dynamical behaviours while they strongly interact with each other? The point is that the dynamical properties of one flavor need exert its influences to another flavor via affecting the gluon self-energy. If the variance of the quark propagator can influence the gluon propagator strongly enough the behaviours of different flavours could be approximately synchronous. But our calculation shows that the reality is in the opposite case. The strongly suppression of the quark loop's contribution to the gluon self-energy by the correction factor $\eta$ in Eq.(\ref{t2}) is helpful to the separation of the $u,d$ quarks' CEPs. But if the influence of quark propagator to the gluon self-energy is too weak, the effect of the chemical potential 'asymmetry' caused by $s$ quark would hardly appear. Because the quarks of different flavors interact directly with each other in the NJL model, the studies based on NJL model can't find such separation \cite{a7}.

\acknowledgments
Thanks for the support of the Nanchong Shi Xiao Ke Ji He Zuo Xiang Mu(NC17YS4014).

\end{document}